\documentclass[runningheads]{llncs}

\usepackage[T1]{fontenc}
\usepackage{graphicx}
\usepackage{amsmath,amssymb}
\usepackage{bbm}
\usepackage{xcolor}
\usepackage{xurl}          
\usepackage{booktabs}
\usepackage{pifont}
\usepackage{IEEEtrantools}
\usepackage[normalem]{ulem}
\usepackage{float}

\begin{document}
\title{Reliable Remediation Impact Prediction for Black-Box Security Ratings}
\titlerunning{
Reliable and Effective Remediation Impact Prediction
}

\author{Nada HANAD\orcidID{0009-0002-2039-6287}
\and
Mehdi Acheli\orcidID{0000-0001-9649-7127} \and
Ali {Nour Eldin}\orcidID{0000-0002-0956-8108} \and
Mohamed Sellami\orcidID{0000-0002-7547-1857}
\and
Walid GAALOUL\orcidID{0000-0003-0451-532X}
}
\authorrunning{N. HANAD et al.}

\institute{SAMOVAR, Telecom SudParis, Institut Polytechnique de Paris, Palaiseau, France\\
\email{nada.hanad@telecom-sudparis.eu}}

\maketitle

\begin{abstract}
Security rating platforms summarize externally observable cyber exposure and are expected to help organizations prioritize remediation. A platform may want to tell an organization how a candidate remediation action would affect its score, but repeatedly exposing exact score responses can reveal information about the hidden scoring engine. We propose a surrogate based approach for remediation score impact prediction that is designed to respect this opacity constraint. The surrogate predicts scores from organization configurations while explicitly representing checkpoint (i.e., a security check) applicability and the observed checkpoint set. A main challenge is that such predictions are not uniformly reliable: they depend on the amount and structure of the observable checkpoint evidence available for a given configuration. To address this, the approach combines applicability-aware surrogate construction, sensitivity analysis under controlled checkpoint restriction, a reliability layer for identifying unstable predictions, and score-impact prediction for supported remediation actions. Explicit modeling of checkpoint applicability is central throughout: it improves score prediction and provides the feature basis used by the reliability layer to identify unstable cases. We evaluate the approach on a real-world dataset of 5,188 organization configurations from a commercial security rating platform. The results show that the applicability-aware surrogate improves score prediction over simpler feature representations. For remediation, the surrogate predicts the score impact of supported actions, while the reliability layer helps identify cases in which these predicted impacts should be interpreted cautiously.
\end{abstract}


\section{Introduction}
\label{sec:intro}

Commercial security rating platforms are increasingly used to summarize cyber exposure from publicly observable evidence such as exposed services, DNS and mail configurations, TLS settings, and publicly reachable web assets. Vendors such as BitSight~\cite{bitsight_platform}, SecurityScorecard~\cite{securityscorecard_platform}, and Board of Cyber~\cite{boardofcyber_securityrating} collect signals from organizations' digital footprint and aggregate them into a numerical score that is then used in procurement, cyber insurance, and third party risk management~\cite{he2024cyber,keskin2021}. In practice, such scores are derived from the outcomes of multiple security checks, which we refer to here as checkpoints. Statistics suggest that these scores are operationally meaningful: BitSight states that companies rated at or below 400 on its 250 to 900 scale are five times more likely to experience a publicly disclosed breach than those rated at or above 700~\cite{bitsight_breach_2022}, while SecurityScorecard reports that organizations receiving an $F$ grade on its $A$ to $F$ scale are 13.8 times more likely to suffer a breach than those with an $A$ grade~\cite{ssc_breach_2022}. This growing operational role of security rating platforms is also reinforced by regulations. In the European Union, for example, the Digital Operational Resilience Act (DORA) requires financial entities to continuously monitor the information and communication technology risk posed by third party providers~\cite{eu_dora_2022}.

Beyond assigning a score, these platforms are also expected to provide useful remediation guidance. In practice, organizations want to know not only their current rating, but also which concrete actions could improve it and by how much. Providing such guidance, however, may leak information about the underlying scoring logic when different remediation options are explored and their predicted score effects can be compared. This is particularly problematic because security rating platforms are designed to remain black boxes, with providers keeping their exact scoring logic proprietary and undisclosed~\cite{keskin2021,woods2021sok}.

In this paper, we address the problem of predicting remediation related score impacts without exposing the underlying scoring engine. Instead of providing organizations with the exact scores corresponding to different candidate post remediation configurations, we propose a machine learning surrogate model that predicts approximate scores, trained on the scoring engine's historical data, and can therefore be used to predict the score impact of candidate remediation actions without exposing exact score responses of the proprietary scoring engine. Because the surrogate only approximates the original scoring engine, its outputs remain predictions rather than direct evaluations of the proprietary function.

However, using a surrogate model to approximate a proprietary scoring engine introduces non uniform deployment-time prediction quality across different organization configurations, meaning that the resulting score predictions are not equally reliable for all configurations. This pattern is particularly visible for lower scoring configurations, whose associated scores are harder to predict. These lower scoring configurations also tend to have fewer applicable checkpoints for evaluation. This suggests that the cases on which the surrogate performs less accurately are not random, and that part of this variation in prediction quality may be driven by the size and structure of the applicable checkpoint set associated with a given organization configuration. This links our problem both to work on informative missingness, where the pattern of what is observed can itself carry predictive information, and to work on predictive reliability, where the goal is to identify when model outputs should be trusted. More broadly, this can be viewed as a machine learning problem of prediction reliability under heterogeneous measurement conditions, in a setting where surrogate score predictions are used to assess candidate remediation actions~\cite{angelopoulos2024crc,lipton2016,vanness2023,sharafoddini2019}.

This observation leads to four research questions. First, if prediction quality varies with the number of applicable checkpoints, how sensitive are deployment-time surrogate score predictions to variations in the number of applicable checkpoints, that is, security checks that are relevant and observable for a given organization configuration? ($RQ1$) Second, what is the impact of explicitly modeling checkpoint applicability on surrogate score prediction? ($RQ2$) Third, can checkpoint applicability identify which organization configurations are prone to unstable predictions, and can formal reliability guarantees be attached to accepted predictions? ($RQ3$) Fourth, can the surrogate reliably predict the score impact of user simulated candidate remediation actions for a given organization configuration? ($RQ4$)

To answer $RQ1$, we study how surrogate predictions change when the observed checkpoint set is deliberately restricted at deployment-time. For $RQ2$, we compare predictions from surrogate representations with and without explicit checkpoint applicability information. For $RQ3$, we build a reliability layer that predicts instability from checkpoint applicability features and attaches selective guarantees using Conformal Risk Control~\cite{angelopoulos2024crc}. For $RQ4$, we use the surrogate to predict the score impact of user simulated candidate remediation actions and compare these predictions against trusted reference scores, while also testing whether predicted instability helps identify cases that should be interpreted more cautiously.

Specifically, we put forward the following contributions:
\begin{itemize}
    \item[(i)] We propose a surrogate based score-prediction approach for black box security rating platforms, designed to assess candidate remediation changes without directly exposing the proprietary scoring engine. As part of this approach, we enrich the surrogate features with checkpoint applicability and related measurement structure.
    
    \item[(ii)] We introduce a reliability layer that uses checkpoint applicability information to identify organization configurations that are prone to unstable surrogate predictions, and we study how formal reliability guarantees can be attached to accepted predictions.

    \item[(iii)] We evaluate the proposed approach along three dimensions: surrogate score prediction and the impact of applicability aware features through an ablation study; prediction sensitivity and reliability under controlled restriction of the observed checkpoint set; and remediation support, by validating predicted score impacts for user simulated candidate remediation actions against trusted reference scores.
\end{itemize}

The rest of the paper is organized as follows. Section~\ref{sec:problem} introduces the motivating example and the problem setting. Section~\ref{sec:rw} reviews related work. Section~\ref{sec:approach} presents the proposed approach. Section~\ref{sec:setup} describes the study setting, evaluation protocol, and experimental results. Section~\ref{sec:discussion} discusses the findings and their limitations. Section~\ref{sec:conclusion} concludes.

\section{Motivating Example}
\label{sec:problem}

\begin{figure}
    \centering
    \includegraphics[width=\linewidth]{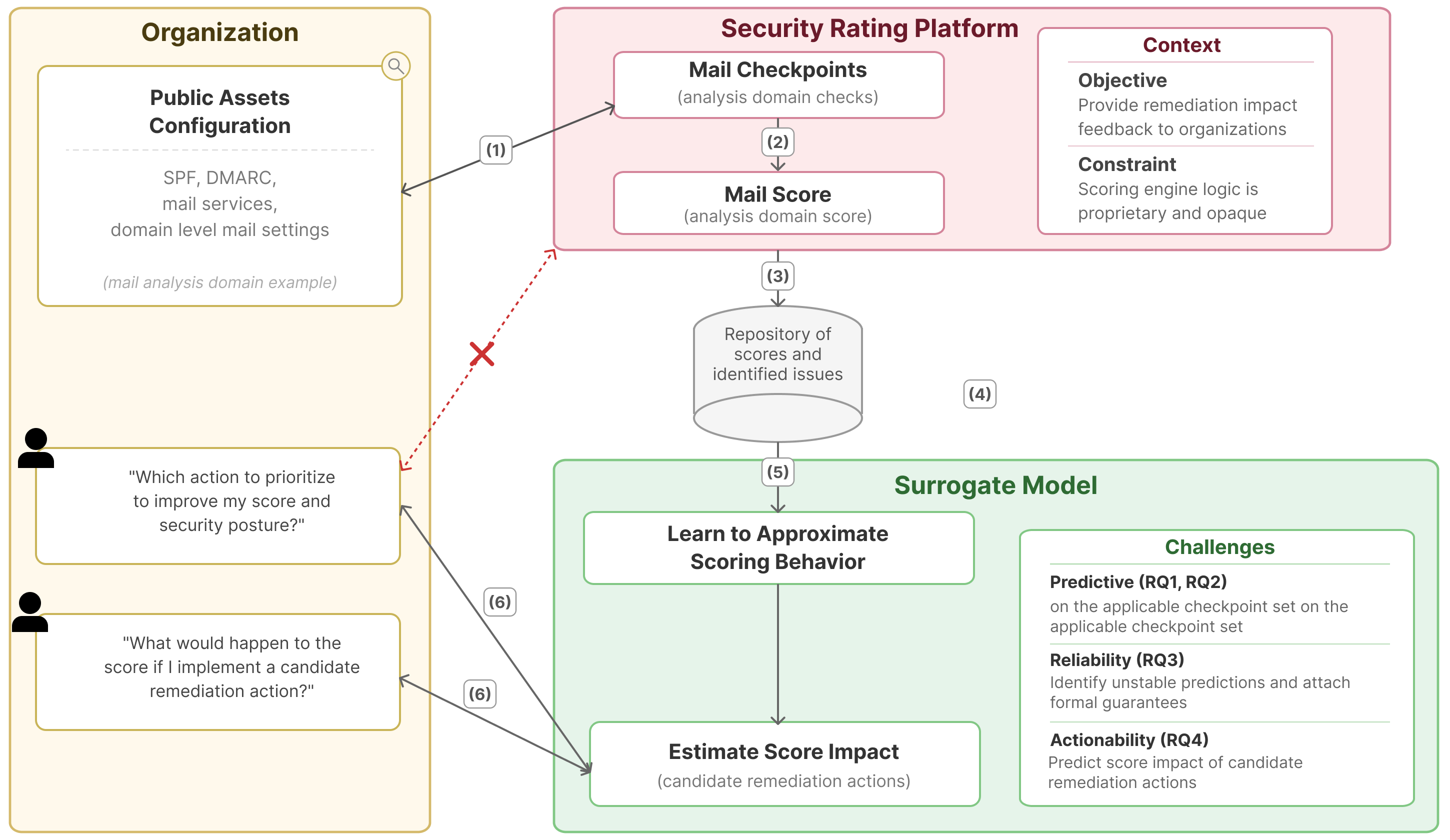}
    \caption{Motivating example for the Mail analysis domain}
    \label{fig:example}
\end{figure}

To motivate our work, we consider a realistic scenario in which an organization monitors and improves its security posture through a commercial security rating platform illustrated in Fig.~\ref{fig:example}. Such platforms typically cover multiple analysis domains (e.g., Attack Surface and Mail) derived from externally observable evidence. For simplicity, however, we restrict the motivating example to the Mail domain. The platform inspects a set of mail-related security checks (see Fig.~\ref{fig:example}, (1)), for example checks associated with mail authentication records, domain-level mail configuration, and exposed mail services, and aggregates the resulting observations into an analysis domain score (see Fig.~\ref{fig:example}, (2)). We call such security checks \emph{checkpoints}. The outcomes of these checkpoints are then processed by the platform’s internal scoring engine to produce a numerical \emph{score} on a 0--100 scale, which is returned to the user alongside a description of the \emph{issues} uncovered by the checkpoints. As shown in Fig.~\ref{fig:example}, these scores and identified issues are then stored in the platform repository (3).

From the organization's perspective, the returned Mail \emph{score} is useful but not sufficient. Knowing this \emph{score} is only part of the problem. The organization also wants to answer a natural what-if question: ``\emph{what would happen to the score if I implemented a candidate remediation action, such as strengthening a mail authentication setting or correcting a misconfigured mail-related service?}'' This corresponds to the lower-left query shown in Fig.~\ref{fig:example}, which is later addressed through the surrogate interaction in step (6). The practical need is therefore not merely to observe the current \emph{score}, but to evaluate the score impact of possible \emph{remediation actions} before deployment.

On one hand, estimating the \emph{score} impact of candidate \emph{remediation actions} is useful for remediation assessment. On the other hand, directly exposing exact \emph{scores} for many candidate \emph{remediation actions} would reveal too much about the proprietary scoring logic. This practical dilemma motivates the use of a \emph{surrogate model}, i.e., a machine learning model trained to approximate the platform’s internal scoring behavior, that predicts scores from an organization configuration while avoiding direct exposure of the underlying scoring engine. In Fig.~\ref{fig:example}, this transition is represented by using the historical scores and issues stored in the repository to learn the surrogate scoring behavior (4), and then using the surrogate to predict score impacts for candidate remediation actions (5).

This setting is complicated by two factors. First, not all checkpoints apply to all organizations. Applicability depends on the technical conditions that make a checkpoint relevant, so two organizations in the same analysis domain may be evaluated through different sets of applicable checkpoints. For example, if an organization does not deploy a given mail security mechanism, then checkpoints that assess properties of that mechanism are inapplicable. Missing checkpoint outcomes are therefore often structural rather than accidental, because a checkpoint may be absent simply because the technical conditions that make it relevant are not satisfied for the assessed organization~\cite{heckman1979,lipton2016,vanness2023,rubin1976,sharafoddini2019}. Second, the scoring engine that maps checkpoint outcomes to a score is proprietary and undisclosed~\cite{keskin2021,woods2021sok}. A platform may display the current score, but it does not reveal the exact function that would map every hypothetical post-remediation configuration to its corresponding score.
The surrogate therefore addresses the original objective shown in Fig.~\ref{fig:example}, namely providing remediation impact feedback despite engine opacity. However, introducing the surrogate does not by itself solve the full problem. It creates a second layer of challenges, also shown in Fig.~\ref{fig:example}, that concern how accurately the surrogate predicts scores, how to detect when its predictions are unstable, and whether its predicted score impacts are reliable enough to support remediation assessment.

\paragraph{\textbf{Problem setting}.} 

We now abstract the motivating example into a more general problem. We consider a black box security rating system that evaluates organizations through analysis domain specific checkpoints and assigns a numerical score from the resulting observations~\cite{keskin2021,woods2021sok}. For a given organization at a given observation time, only a subset of the full checkpoint set may be applicable. We refer to the resulting analysis domain specific observation, consisting of the applicable checkpoints and their observed outcomes, as an \emph{organization configuration}.

The first challenge is predictive. If a surrogate model is used to predict scores from organization configurations, average predictive accuracy alone is not sufficient. Prediction quality may vary across configurations because different configurations are associated with different sets of applicable checkpoints. This is the predictive challenge highlighted in Fig.~\ref{fig:example}, and is the basis of $RQ1$ and $RQ2$. This motivates our analysis of sensitivity to checkpoint availability and whether explicit modeling of checkpoint applicability improves surrogate score prediction.

The second challenge is reliability. If prediction quality is not uniform across configurations, then a system needs a way to identify configurations that are more likely to yield unstable surrogate predictions. This corresponds to the reliability challenge in Fig.~\ref{fig:example}, which motivates $RQ3$. This motivates the design of a reliability layer based on checkpoint applicability information, together with formal reliability guarantees attached to accepted predictions~\cite{angelopoulos2024crc}.

The third challenge is actionability. The goal is not for the model to invent remediation actions on its own. Rather, a user specifies a candidate remediation action, the corresponding post-change configuration is constructed, and the surrogate is used to predict the resulting score and thus the impact associated with that action. In this setting, the key empirical question is whether these predicted score impacts are reliable enough to support remediation assessment in practice.
This is the actionability challenge in Fig.~\ref{fig:example}, and it motivates $RQ4$.

\section{Related Work}
\label{sec:rw}

This section reviews related work on commercial security rating systems and external cyber measurement (Section~\ref{sec:rw:sr}), on missing data and informative missingness in predictive models (Section~\ref{sec:rw:missing}), and on reliability in predictive models (Section~\ref{sec:rw:reliability}).

\subsection{Security Rating Systems}\label{sec:rw:sr}

Commercial security rating systems are widely used in practice, but their meaning and validity remain debated~\cite{keskin2021,woods2021sok}. Keskin et al.~\cite{keskin2021} compared several non intrusive scoring platforms and found substantial disagreement across vendors evaluating the same organizations. Their results raise a basic interpretability question: when two platforms observe the same public facing surface and disagree, it becomes unclear what exactly each score is measuring. Woods and B{\"o}hme~\cite{woods2021sok} reached a related conclusion at a broader level in their systematization of cyber risk quantification. They showed that observable signals can be useful, but that the link between such signals and the broader notion of cyber risk is indirect and depends strongly on the design of the measurement process.

Earlier measurement work also showed that organization level security assessment is feasible, but highly sensitive to modeling choices. Noroozian et al.~\cite{noroozian2017} combined heterogeneous abuse datasets and used Item Response Theory to estimate provider security performance as a latent trait. Tajalizadehkhoob et al.~\cite{tajalizadehkhoob2017} similarly showed that conclusions about responsibility for weak security depend strongly on how the measurement problem is structured. Taken together, these studies support the idea that external cyber measurement can provide useful signals, while showing that its interpretation depends on the scoring design itself.

In the broader literature on black-box model auditing, surrogate models have been used to analyze opaque predictive systems. Tan et al.~\cite{tan2018distill}, for example, use transparent student models to mimic black-box risk scores for auditing. However, their focus is model auditing rather than surrogate reliability under varying measurement conditions, and they do not study score changes under hypothetical interventions.

Overall, these works focus on validity, interpretation, or disagreement across external cyber scores, or on auditing opaque risk scores through surrogate distillation. They do not address how surrogate prediction behavior varies with measurement structure, such as checkpoint applicability and the size of the observed checkpoint set associated with a given organization configuration, nor whether such a surrogate can predict score changes under hypothetical interventions such as user simulated candidate remediation changes.

\subsection{Missing Data and Informative Missingness in Predictive Models}\label{sec:rw:missing}

Missingness refers to the absence of values that could in principle have been observed in the data. Rubin~\cite{rubin1976} introduced the classical taxonomy of missingness mechanisms, and Heckman~\cite{heckman1979} showed that the process determining what is observed can itself affect inference. In predictive settings, Lipton et al.~\cite{lipton2016} showed that modeling missingness patterns explicitly can improve clinical time series classification. Sharafoddini et al.~\cite{sharafoddini2019} similarly highlighted that missingness in intensive care data can themselves be informative rather than reflecting random noise. Van Ness et al.~\cite{vanness2023} further showed that missing indicators can remain useful predictive signals even in higher dimensional settings.

These works are directly relevant to our problem because they show that the pattern of what is observed can itself carry information. In our case, missing checkpoint outcomes are structural rather than random: a value is absent because its checkpoint is not applicable. Checkpoint applicability therefore forms part of the observed measurement basis itself. Existing work on missingness supports the idea that such structural information may carry predictive signal, but it has not been studied in the context of black-box security rating systems.

A narrower technical connection can be found in work on feature decrement and missingness shift. Cheng et al.~\cite{cheng2025feature} study settings where information available during training becomes unavailable at test time, and Rockenschaub et al.~\cite{rockenschaub2024} study prediction under changes in the missingness process. These settings are not the main framing of our work, since low checkpoint number is already present in the training data rather than arising only at deployment time. However, they remain relevant to the controlled checkpoint restriction analysis that we use later as a diagnostic sensitivity test.
\subsection{Reliability in Predictive Models}

\label{sec:rw:reliability}
Work on reliability in predictive models has developed along several closely related lines. Chow~\cite{chow1970} introduced the classical reject option formulation, where a model may abstain rather than make a low confidence prediction. El-Yaniv and Wiener~\cite{elyaniv2010} studied the foundations of selective classification, and Geifman and El-Yaniv~\cite{geifman2017,geifman2019} extended this perspective to modern neural predictors. Taken together, these works study how predictive systems can identify cases on which they are less likely to be reliable.

A related line of work studies reliability guarantees through conformal methods. Angelopoulos et al.~\cite{angelopoulos2024crc} introduced Conformal Risk Control, which allows guarantees to be attached to accepted predictions with respect to a user defined risk criterion rather than accuracy alone.

These works are directly relevant to our reliability layer. Nevertheless, prior work has not connected selective prediction or conformal-style reliability guarantees to the reliability of surrogate predictions under varying measurement conditions, nor to the identification of configurations that are prone to instability when the available evidence is limited.

\textbf{Discussion.} Existing work addresses important parts of our problem, but in separate settings and for different objectives. We draw on the security rating literature for the broader context of black-box scoring systems, on the missingness literature to motivate checkpoint applicability as informative signal, and on the reliability literature, especially selective prediction and conformal risk control, to design a mechanism for identifying and selectively accepting predictions that are less likely to be unstable. Our contribution is to connect these strands in a single security rating setting through a surrogate-based approach that models checkpoint applicability, studies prediction sensitivity to available checkpoint evidence, adds a reliability layer with formal guarantees, and evaluates remediation score impact prediction without exposing the proprietary scoring function.

\section{Applicability-Aware and Reliable Surrogate Prediction}
\label{sec:approach}

\begin{figure}
    \centering
   \includegraphics[width=1.1\linewidth]{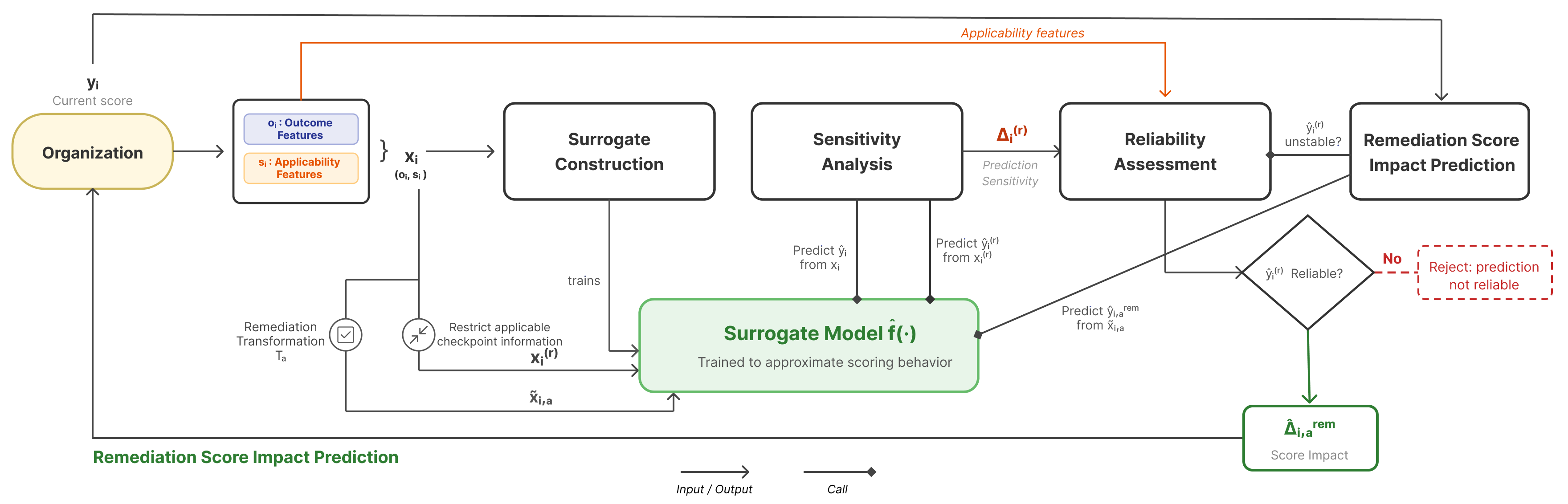}
\caption{Overview of the proposed approach.}
    \label{fig:approach}
\end{figure}

Our approach consists of four linked phases as illustrated in Fig.~\ref{fig:approach}. We first construct a surrogate model that predicts analysis domain scores from organization configurations while explicitly encoding checkpoint applicability. We then apply controlled restriction of the applicable checkpoint information as a diagnostic sensitivity test in order to study when surrogate predictions become unstable. On top of this diagnostic, we build a reliability layer that uses checkpoint applicability information to identify configurations whose surrogate predictions are more likely to be unstable, and to define a selective acceptance rule with finite sample risk guarantees. Finally, we use the surrogate to predict the score impact of user simulated candidate remediation actions.

Figure~\ref{fig:approach} summarizes how these four phases are connected. Starting from an organization configuration, represented by outcome features $o_i$ and applicability features $s_i$, the surrogate model produces a baseline score prediction. A restricted version of the same representation is then used in the sensitivity phase to measure score drift under reduced applicable checkpoint information. This drift defines the instability labels used by the reliability phase, while the original representation is also transformed under a candidate remediation action to estimate the corresponding score impact. Together, the four phases define the path from an observed organization configuration to a score prediction, an instability assessment, and a remediation score impact prediction.

In the following sections we detail these four phases: surrogate construction in Section~\ref{subsec:surrogate}, sensitivity analysis in Section~\ref{subsec:sensitivity}, reliability assessment in Section~\ref{subsec:reliability}, remediation score impact prediction in Section~\ref{subsec:remediation}.

\subsection{Surrogate Construction}
\label{subsec:surrogate}

In this phase, we take as input an organization configuration and construct both its feature representation and the surrogate model used throughout the rest of the approach.
Let $\mathcal{C}=\{1,\dots,M\}$ denote the full set of checkpoints defined by a security rating platform provider for an analysis domain $\mathcal{D}$. For an organization's configuration $i$, each checkpoint $j \in \mathcal{C}$ is either applicable or not. We denote checkpoint applicability by $a_{ij} \in \{0,1\}$, where $a_{ij}=1$ means that checkpoint $j$ applies to configuration $i$ and can therefore contribute an observed outcome, while $a_{ij}=0$ means that the checkpoint does not apply. When a checkpoint $j$ is applicable, we denote its observed outcome by $z_{ij}$.

The observed checkpoint set for a configuration $i$ is therefore defined as
\begin{IEEEeqnarray}{rCl}
\mathcal{C}_i = \{j \in \mathcal{C} : a_{ij}=1\},
\end{IEEEeqnarray}
and the corresponding number of applicable checkpoints is
\begin{IEEEeqnarray}{rCl}
K_i = \sum_{j=1}^{M} a_{ij}.
\end{IEEEeqnarray}

Next, we construct a feature representation $x_i=(o_i,s_i)$ for each organization configuration $i$. The block $o_i$ summarizes checkpoint outcomes observed on applicable checkpoints, while $s_i$ summarizes checkpoint applicability and the structure of the applicable checkpoint set. This design lets the surrogate distinguish weak security evidence from limited observed evidence: missing or sparse values in $o_i$ are interpreted together with explicit information in $s_i$ about which checkpoints apply and how much checkpoint evidence is available.

The surrogate model is defined as a predictor
\begin{IEEEeqnarray}{rCl}
\hat{f}: x_i \mapsto \hat{y}_i,
\end{IEEEeqnarray}
where $\hat{y}_i$ is the predicted analysis domain score for organization configuration $i$. This predictor underlies the rest of the approach: it provides the baseline predictions for the sensitivity analysis, defines the predictions whose reliability is assessed later, and is used to estimate the score impact of user-simulated candidate remediation actions.

In later sections, specifically in Section~\ref{sec:setup:surrogate_restriction}, we evaluate the applicability-aware representation against simpler alternatives that rely only on checkpoint outcome information, in order to determine whether explicit modeling of checkpoint applicability improves surrogate score prediction. This phase therefore provides the model construction used to study the predictive challenge associated with checkpoint applicability, which is evaluated in the context of $RQ2$. We now turn to the second phase, which studies how this surrogate behaves when part of the applicable checkpoint information is removed. 

\subsection{Sensitivity Analysis}
\label{subsec:sensitivity}

We use controlled restriction of applicable checkpoints as a diagnostic sensitivity test. The purpose of this test is to examine whether the score predictions produced by the surrogate model constructed in Section~\ref{subsec:surrogate} are sensitive to changes in the checkpoint information available to the model. Starting from the feature representation $x_i$ of an original organization configuration, we define a restricted representation
\begin{IEEEeqnarray}{rCl}
x_i^{(r)} = R_r(x_i),
\end{IEEEeqnarray}
where $R_r$ uses the subscript $r$ to index the restriction policy being applied. It is a restriction operator that hides part of the applicable checkpoint information according to a predefined restriction policy while keeping the underlying organization configuration fixed. Intuitively, $x_i^{(r)}$ represents the same organization configuration evaluated from a reduced set of available checkpoint information, not a different underlying configuration. In our setting, the restriction operator therefore masks part of the applicable checkpoint information and recomputes any derived features consistently. The surrogate prediction for this restricted representation is
\begin{IEEEeqnarray}{rCl}
\hat{y}_i^{(r)} = \hat{f}(x_i^{(r)}).
\end{IEEEeqnarray}

We quantify prediction sensitivity under restriction through the induced score change: 
\begin{IEEEeqnarray}{rCl}
\Delta_i^{(r)} = \hat{y}_i^{(r)} - \hat{y}_i.
\end{IEEEeqnarray}
This quantity provides a direct diagnostic of fragility under restriction and the basis for defining instability in the next phase.

For a given threshold $\tau > 0$, we say that organization configuration $i$ yields an \emph{unstable prediction} under restriction if
\begin{IEEEeqnarray}{rCl}
|\Delta_i^{(r)}| > \tau.
\end{IEEEeqnarray}

For later use for the reliability assessment, we define the corresponding unstable-versus-stable binary instability label: 
\begin{IEEEeqnarray}{rCl}
\label{eq:binlab}
u_i^{(r,\tau)} = \mathbf{1}\!\left(|\Delta_i^{(r)}| > \tau\right).
\end{IEEEeqnarray}
This label is obtained by thresholding the score drift induced by the restriction procedure and provides the supervision target for the instability predictor introduced in Section~\ref{subsec:reliability}.

This diagnostic allows us to study, in a controlled way, whether surrogate predictions become fragile when the available applicable checkpoint information is reduced. It also gives an operational definition of instability that can later be used in the reliability assessment phase. This phase therefore addresses the predictive challenge identified in Section~\ref{sec:problem} and provides the protocol for $RQ1$, which asks whether surrogate score predictions are sensitive to variation in checkpoint number. We next use this definition of instability to train a reliability model that predicts when surrogate outputs should be trusted.

\subsection{Reliability Assessment}
\label{subsec:reliability}

This phase assesses whether a given organization configuration is likely to yield an unstable surrogate prediction and whether predictions can be selectively accepted under a controlled instability risk. Here, instability risk is defined with respect to the diagnostic restriction in Section~\ref{subsec:sensitivity}. Let $s_i$ denote the applicability and measurement-structure feature block used by the reliability layer. The layer includes a reliability model defined by
\begin{IEEEeqnarray}{rCl}
g: s_i \mapsto p_i^{\mathrm{unstable}}
\end{IEEEeqnarray}
where $p_i^{\mathrm{unstable}}$ is the predicted probability that configuration $i$ yields an unstable surrogate prediction under the diagnostic restriction.

Operationally, this phase has three parts. First, the sensitivity analysis defines the binary instability label $u_i^{(r,\tau)}$ in Eq.~\ref{eq:binlab}. Second, the model $g$ is trained on $s_i$ to predict this label and output $p_i^{\mathrm{unstable}}$. Third, this probability is converted into an acceptance rule that retains only predictions whose instability risk is sufficiently low for subsequent score prediction and remediation analysis.

The reliability layer serves two purposes. First, it uses $s_i$ to identify organization configurations whose surrogate predictions are more likely to be unstable. Second, it provides a basis for selectively accepting predictions under finite-sample risk guarantees. To do so, we apply Conformal Risk Control (CRC)~\cite{angelopoulos2024crc}. We use a held-out calibration set to define the acceptance rule. The calibration and future configurations to which this rule is applied are assumed exchangeable, which is the standard assumption used by conformal methods~\cite{barber2023beyond}.

Concretely, let $\lambda \in [0,1]$ be a threshold on the predicted instability probability. A configuration is accepted when its predicted instability probability is no larger than $\lambda$:
\begin{IEEEeqnarray}{rCl}
A_{\lambda} = \{ i : p_i^{\mathrm{unstable}} \le \lambda \},
\end{IEEEeqnarray}
The corresponding selective instability risk is
\begin{IEEEeqnarray}{rCl}
\mathcal{R}(\lambda) = \mathbb{E}\!\left[u_i^{(r,\tau)} \mid i \in A_{\lambda}\right].
\end{IEEEeqnarray}

Given a target risk level $\alpha$, CRC selects an acceptance threshold $\lambda$ such that the instability risk among accepted predictions remains below $\alpha$. We denote the retained threshold by $\lambda^\star$ and the corresponding accepted set by $A_{\lambda^\star}$. If no feasible threshold is found, then $A_{\lambda^\star}=\emptyset$. Under the exchangeability assumption stated above, this yields a finite-sample guarantee with confidence at least $1-\delta$ that the instability risk among accepted predictions does not exceed $\alpha$.

The practical role of this phase is to determine when the surrogate's score predictions should be trusted, thereby addressing the reliability challenge identified in Section~\ref{sec:problem} and providing the method used for $RQ3$.

\subsection{Remediation Score Impact Prediction}
\label{subsec:remediation}

The final phase focuses on remediation score impact prediction. Rather than generating remediation actions automatically, the surrogate model takes a user-defined candidate action for a given organization configuration and predicts the resulting score and score impact. We also use the reliability layer to accept such score impacts as dependable. In our experiments, these actions correspond to structured interventions such as deploying SPF or deploying DMARC (Mail analysis domain).

For a configuration $i$ considered for remediation, the input to this phase is the observed representation $x_i$ together with a user-specified candidate action $a$. The output is a post-action score prediction, the corresponding predicted score impact relative to the current score and whether this impact is accepted.

Let $\mathcal{I}$ denote the set of organization configurations considered for remediation. For an organization's configuration $i \in \mathcal{I}$, $y_i$ denotes its current score. The score impact is defined relative to $y_i$. Let $T_a$ denote the transformation induced by candidate remediation action $a$. Applying $T_a$ produces a post-action organization configuration $i'_a$, with the representation:
\begin{IEEEeqnarray}{rCl}
x_{i'_a} = T_a(x_i).
\end{IEEEeqnarray}
The surrogate then predicts the score of $i'_a$ as
\begin{IEEEeqnarray}{rCl}
\hat{y}_{i'_a} = \hat{f}(x_{i'_a}),
\end{IEEEeqnarray}
and the predicted score impact of $a$ is
\begin{IEEEeqnarray}{rCl}
\widehat{\Delta}_{i,a}^{\mathrm{rem}} = \hat{y}_{i'_a} - y_i.
\end{IEEEeqnarray}

We rely on the reliability layer (Section~\ref{subsec:reliability}) to score the post-action configuration. Given $i'_a$, we extract its checkpoint applicability features $s_{i'_a}$ and compute
\begin{IEEEeqnarray}{rCl}
p_{i'_a}^{\mathrm{unstable}} = g(s_{i'_a}).
\end{IEEEeqnarray}
Exploiting the calibrated acceptance threshold $\lambda^\star$ from Section~\ref{subsec:reliability}, we accept the score as dependable if
\begin{IEEEeqnarray}{rCl}
p_{i'_a}^{\mathrm{unstable}} \le \lambda^\star.
\end{IEEEeqnarray}
Thus, a remediation prediction is retained only when the predicted instability probability of its post-action configuration does not exceed $\lambda^\star$. 


An important requirement is that $T_a$ must produce a semantically valid post-action representation in surrogate feature space. In particular, when a candidate action changes the technical conditions that determine checkpoint applicability, the relevant checkpoint outcomes, the applicable checkpoint set, and all derived features must be updated consistently.

Under this formulation, the surrogate does not recommend actions and does not optimize over possible remediations. Its role is narrower: given an observed organization configuration and a user-simulated candidate action, it predicts the resulting score and the corresponding score impact. The reliability layer then filters non dependable actions.

This phase therefore addresses the actionability challenge identified in Section~\ref{sec:problem} and provides the method used for $RQ4$. 

\section{Experimental Setup and Evaluation}
\label{sec:evaluation}
 \label{sec:setup}

This section evaluates the proposed approach on the Mail analysis domain. We first describe the study setting and protocols in Section~\ref{sec:setup:data_protocol}, then report results for surrogate construction, sensitivity analysis, reliability assessment, and remediation score impact prediction.

\subsection{Study Setting, Data, and Evaluation Protocol}
\label{sec:setup:data_protocol}

We instantiate the problem on the Mail analysis domain of one commercial security rating platform. The archive contains timestamped records that combine checkpoint outcomes with externally observable assets such as domains, URLs, IP addresses, and mail related services. Checkpoints are evaluated only when relevant to those assets, and each detected issue is assigned one of five severity levels, e.g., "Low" or "High".

Because the same checkpoint configuration can appear repeatedly, either over time for one organization or across different organizations, we deduplicate observations across checkpoint related fields while ignoring organization identifier and timestamp. This yields 5,188 distinct organization configurations from 375,408 original records.

For each retained configuration, the dataset includes the platform assigned Mail score and 40 Mail checkpoints. Each checkpoint contributes outcome features describing the number of measured assets, the worst observed severity, and the count and proportion of each severity level, for a total of 12 outcome fields per checkpoint. We then add one applicability indicator per checkpoint together with structural features summarizing the observed checkpoint set and its measurement pattern, including the number and fraction of applicable checkpoints, summaries of applicability distribution, and configuration level missingness and severity summaries. The final dataset contains 533 columns in total.

Our evaluation follows the four phases of Section~\ref{sec:approach}. Table~\ref{tab:protocol_overview} summarizes the protocols. Whenever a split (train/val/test or train/test) is used, it is grouped by organization identifier so that configurations from the same identifier never appear in different partitions of the same experiment. The Repetitions column indicates how many times each split and subsequent experiment is performed.

\begin{table}[t]
\centering
\caption{Evaluation protocols used in the paper.}
\label{tab:protocol_overview}
\footnotesize
\begin{tabular}{p{0.29\linewidth}p{0.18\linewidth}p{0.22\linewidth}p{0.19\linewidth}}
\toprule
\textbf{Evaluation phase} & \textbf{Research questions} & \textbf{Protocol} & \textbf{Repetitions} \\
\midrule
Surrogate Construction & $RQ2$ & 80/10/10 & 20 seeds \\
Sensitivity Analysis & $RQ1$ & 80/20 & 20 seeds \\
Reliability Assessment & $RQ3$ & 60/20/20 & one seed \\
Remediation Score Impact Prediction & $RQ4$ & No split & -- \\
\bottomrule
\end{tabular}
\end{table}

For the sensitivity analysis, the source dataset on which the split is applied is a 10\% held-out test partition from the previous phase. For reliability assessment too, we consider a held out 20\% partition test from the previous experiment and apply the split on it. Since there is no training notion for the fourth phase, no split is applied.

\subsection{Surrogate Construction and Score Prediction}
\label{sec:setup:surrogate_restriction}
\label{sec:results:rq2}
The enriched applicability aware surrogate is a LightGBM regressor~\cite{ke2017lightgbm} trained to predict the platform assigned Mail score. It uses 500 trees, learning rate 0.03, 63 leaves, row subsampling 0.8, and feature subsampling 0.8, and is reused in the later sensitivity, reliability, and remediation phases. For each grouped seed, the model is trained on the training partition, validation MAE is measured on the 10\% validation split, and comparison is based on the held out 10\% test split.

We compare this model with two simpler baselines: an outcome only representation without explicit applicability or structural features, and an applicability only ablation. Table~\ref{tab:feature_sets} summarizes the three representations.

\begin{table}[t]
\centering
\caption{Feature representations used for the first experiment.}
\label{tab:feature_sets}
\footnotesize
\begin{tabular}{p{0.24\linewidth}p{0.51\linewidth}r}
\toprule
\textbf{Representation} & \textbf{Contents} & \textbf{Input fields} \\
\midrule
Outcome & Outcome features & 480 \\
Applicability & 40 checkpoint applicability indicators plus engineered structural fields & 53 \\
Full & Outcome features, checkpoint applicability indicators, and engineered structural fields & 533 \\
\bottomrule
\end{tabular}
\end{table}

The applicability features encode which checkpoints apply and how much checkpoint evidence is available. Organization identifier and timestamp are excluded from the inputs.

Table~\ref{tab:rq2_ablation} shows a clear separation between the full model and the two simpler baselines. The full representation achieves the lowest test error at $0.7816 \pm 0.0327$, versus $1.4554 \pm 0.1565$ for the outcome only baseline and $4.6388 \pm 0.1972$ for the applicability only baseline.

These results show that explicit applicability and measurement structure features materially improve score prediction when combined with checkpoint outcome features. The applicability only ablation also confirms that applicability information alone is not sufficient.

\begin{table}[H]
\centering
\caption{Results of the ablation study.}
\label{tab:rq2_ablation}
\footnotesize
\begin{tabular}{lcc}
\toprule
\textbf{Representation} & \textbf{Validation MAE} & \textbf{Test MAE} \\
\midrule
Full & 0.8154 $\pm$ 0.0745 & 0.7816 $\pm$ 0.0327 \\
Outcome-only & 1.4935 $\pm$ 0.1836 & 1.4554 $\pm$ 0.1565 \\
Applicability-only & 4.6872 $\pm$ 0.1258 & 4.6388 $\pm$ 0.1972 \\
\bottomrule
\end{tabular}
\end{table}

\subsection{Sensitivity Analysis}
\label{sec:results:rq1}

To answer $RQ1$, we apply controlled checkpoint restriction to the full surrogate. For each seed, checkpoints are ranked by applicability on the 80\% training partition, and only the top $K$ checkpoints are retained when evaluating restriction statistics on the held out 20\% test partition:
\[
K \in \{40, 35, 30, 25, 22, 18, 14, 10\}.
\]
For checkpoints not retained at a given $K$, the applicability indicator is set to 0 and all associated outcome fields are set to missing. Engineered structural fields are then recomputed to keep the restricted representation internally consistent.

For each configuration $i$, we compare unrestricted and restricted surrogate predictions. We report $\Delta_i^{(K)}$, the mean induced score change, the 90th percentile of $|\Delta_i^{(K)}|$, and the fraction of cases with $|\Delta_i^{(K)}|>2$.

Table~\ref{tab:rq1_restriction} shows that the surrogate remains unchanged at $K=40$ and $K=35$, and still essentially unchanged at $K=30$. Drift stays limited at $K=25$ and $K=22$, where the 90th percentile reaches only $0.5145$ and $0.7862$, and the fraction of cases with $|\Delta_i^{(K)}| > 2$ remains below $2.3\%$.

The main transition occurs between $K=22$ and $K=18$. The 90th percentile of the absolute score change rises from $0.7862$ to $10.6613$, and the fraction of unstable predictions rises from $2.27\%$ to $69.38\%$. Below this regime, instability remains high, reaching $83.50\%$ at $K=14$ and $97.43\%$ at $K=10$.

We therefore use the transition between $K=22$ and $K=18$ as the main diagnostic regime for the rest of the evaluation.

\begin{table}[H]
\centering
\caption{Surrogate sensitivity under controlled checkpoint restriction. Values are means across seeds with standard deviations in parentheses.}
\label{tab:rq1_restriction}
\footnotesize
\begin{tabular}{lccc}
\toprule
\textbf{$K$} & \textbf{Mean $\Delta_i^{(K)}$} & \textbf{$p90(|\Delta_i^{(K)}|)$} & \textbf{$\%(|\Delta_i^{(K)}| > 2)$} \\
\midrule
40 & 0.0000 (0.0000) & 0.0000 (0.0000) & 0.00 (0.00) \\
35 & 0.0000 (0.0000) & 0.0000 (0.0000) & 0.00 (0.00) \\
30 & 0.0084 (0.0041) & 0.0000 (0.0002) & 0.00 (0.00) \\
25 & -0.0931 (0.0267) & 0.5145 (0.0738) & 1.43 (0.56) \\
22 & -0.1355 (0.0251) & 0.7862 (0.0860) & 2.27 (0.62) \\
18 & 4.4723 (0.6027) & 10.6613 (0.9142) & 69.38 (6.63) \\
14 & 8.0524 (0.9394) & 15.8331 (1.2710) & 83.50 (4.17) \\
10 & 20.2486 (1.2704) & 35.2518 (2.3442) & 97.43 (0.50) \\
\bottomrule
\end{tabular}
\end{table}

\subsection{Reliability Assessment and CRC}
\label{sec:setup:reliability}

In the third experiment, instability labels are derived from the $K=18$ diagnostic restriction. A configuration is labeled unstable when the absolute score change exceeds $\tau$; the main setting uses $\tau=2$, with additional sweeps over $\tau \in \{1,2,3\}$.

The reliability model from Section~\ref{subsec:reliability} is a LightGBM classifier using checkpoint applicability features. It is trained on the derived training split, calibrated on the validation split for CRC threshold selection, and evaluated on the held out test split.

For selective reliability control, we use CRC~\cite{angelopoulos2024crc} on the validation partition. We search over thresholds on predicted instability probability and retain the largest accepted set whose one sided Clopper--Pearson upper bound on accepted instability rate is at most $\alpha$~\cite{clopper1934use}. We evaluate $\alpha \in \{0.05, 0.10, 0.15, 0.20, 0.25, 0.30\}$ with confidence parameter 0.05, then apply feasible thresholds to the test partition.

We report ROC AUC~\cite{fawcett2006introduction}, average precision~\cite{davis2006relationship}, Brier score~\cite{brier1950verification}, accepted coverage, and observed instability rate among accepted test cases.

\label{sec:results:rq3}
We now report results for instability prediction and CRC selective acceptance.

Table~\ref{tab:rq3_instability} reports instability prediction results across the three thresholds. At $\tau=1$, the model reaches ROC AUC $0.8872$ and Brier score $0.1406$. At the main threshold $\tau=2$, performance improves to ROC AUC $0.9255$ and Brier score $0.1001$. At $\tau=3$, ranking remains strong, with ROC AUC $0.9240$ and Brier score $0.0723$.

These results show that checkpoint applicability and measurement structure features are informative for identifying unstable predictions under the diagnostic restriction regime, with the strongest ranking performance at $\tau \in \{2,3\}$.

\begin{table}[t]
\centering
\caption{Instability prediction results for the reliability model across thresholds.}
\label{tab:rq3_instability}
\footnotesize
\begin{tabular}{ccccc}
\toprule
\textbf{$\tau$} & \textbf{ROC AUC} & \textbf{AP} & \textbf{Brier} & \textbf{AURC} \\
\midrule
1 & 0.8872 & 0.8995 & 0.1406 & 0.2299 \\
2 & 0.9255 & 0.8639 & 0.1001 & 0.0721 \\
3 & 0.9240 & 0.6432 & 0.0723 & 0.0210 \\
\bottomrule
\end{tabular}
\end{table}

CRC shows that selective guarantees are feasible, with achievable coverage depending on $\tau$ and $\alpha$. For $\tau=1$, no feasible threshold is found below $\alpha=0.15$, and accepted test coverage rises from $23.1\%$ at $\alpha=0.15$ to $38.5\%$ at $\alpha=0.30$. For the main threshold $\tau=2$, feasibility also begins at $\alpha=0.15$, with coverage increasing from $16.3\%$ to $89.4\%$ as $\alpha$ is relaxed to $0.30$. For $\tau=3$, feasibility starts at $\alpha=0.10$, with $78.8\%$ accepted coverage, and reaches $98.1\%$ for $\alpha \geq 0.15$. Overall, the pipeline supports practical selective guarantees at substantial coverage for moderate and larger instability thresholds.

\subsection{Remediation Score Impact Prediction}
\label{sec:setup:remediation_metrics}

To answer $RQ4$, we evaluate two candidate remediation action families in the Mail analysis domain: deploying SPF and deploying DMARC. For each action family, source pools are built from observed configurations that already exhibit the corresponding post-action pattern and serve as a reference for the transformation. Post-action configurations are then generated for each eligible configuration (configurations for which the action is plausible in real-world settings) and applicability summaries and structural fields are recomputed for each post-action configuration. Practically, we evaluate predicted score impacts against trusted reference gains. We also examine whether predicted instability distinguishes more dependable remediation predictions from less dependable ones.

\label{sec:results:rq4}
Table~\ref{tab:rq4_validation} shows that, on eligible cases, both targeted actions produce positive average predicted gains. This is realistic since the actions are remediative in nature and contain fixes that should improve the score. SPF deployment applies to 8 eligible configurations while DMARC deployment applies to 51. The MAE column indicates a moderate mean absolute error between the predicted gains and reference ones with a stronger performance for DMARC than for SPF. We notice a divergence in these MAE compared with the initial test MAE reported for the enriched surrogate model in subsection \ref{sec:results:rq2}. 



\begin{table}[t]
\centering
\caption{Remediation score impact summaries and validation against reference gains.}
\label{tab:rq4_validation}
\footnotesize
\begin{tabular}{lccc}
\toprule
\textbf{Action family} & \textbf{Eligible} & \textbf{          Mean Predicted Gain          } & \textbf{MAE} \\
\midrule
Deploy SPF & 8 & 10.11 & 15.00 \\
Deploy DMARC & 51 & 9.70 & 8.20 \\
\bottomrule
\end{tabular}
\end{table}

As a solution, we use our reliability layer to help interpret these remediation predictions. We perform a reliability analysis on DMARC post-action cases, split by predicted instability probability. Using a threshold of $\lambda^\star = 0.5$, $22.6\%$ of the cases fall into the higher instability non-accepted group and $77.4\%$ into the lower instability accepted group. The non-accepted group has mean predicted gain $-1.58$, with $71.4\%$ negative score impacts and $9.35$ MAE. By contrast, the accepted group has a mean predicted gain of $11.26$, no negative impact scores and $6.94$ MAE.

This suggests that the reliability layer adds value by separating cases where predicted score gains are more dependable from cases that should be interpreted more cautiously. Importantly, cases where the predicted gain is far from the trusted reference gain tend also to be flagged as unstable. Large remediation errors therefore do not appear as unexplained failures of the surrogate, but as part of a pattern already captured by the instability signal. This provides a useful warning signal for cases in which the predicted gain should not be trusted at face value.

Table~\ref{tab:rq4_cases} illustrates this pattern with four anonymized DMARC cases. Cases A--C have high predicted instability and predicted gains near or below zero even though the reference gains are highly positive. Case D has low predicted instability, a large positive predicted gain close to the reference gain. These examples show at the case level that mismatched remediation estimates tend to coincide with high predicted instability, whereas the most accurate and clearly positive estimates appear in low instability cases.

\begin{table}[H]
\centering
\caption{Illustrative DMARC cases for remediation oriented reliability analysis. The reference gain is reported for comparison.}
\label{tab:rq4_cases}
\scriptsize
\begin{tabular}{lcccc}
\toprule
\textbf{Illustrative case} & \textbf{Instab. prob.} & \textbf{Predicted Impact} & \textbf{Reference gain} & \textbf{MAE} \\
\midrule
A & 0.823 & -1.82 & 25.00 & 26.82\\
B & 0.766 & -2.65 & 25.00 & 27.65\\
C & 0.766 & -1.45 & 25.00 & 26.45\\
D & 0.154 & 22.83 & 30.00 & 7.17 \\
\bottomrule
\end{tabular}
\end{table}

Overall, the $RQ4$ results show that surrogate based remediation score impact prediction is feasible for the evaluated action families. SPF and DMARC deployment produce positive predicted gains on eligible cases, with lower error for DMARC than for SPF. The remediation oriented reliability analysis further shows that cases with large discrepancies from trusted reference gains are typically the same cases flagged as unstable, which makes the reliability layer useful for identifying predictions that should be interpreted cautiously.

 \section{Discussion and Future Work}
 \label{sec:discussion}

The results show that checkpoint applicability matters for both score prediction and reliability. When modeled jointly with checkpoint outcomes, it improves surrogate performance over both the outcome only and applicability only baselines. It is also informative for reliability assessment: under controlled checkpoint restriction, the surrogate remains stable when enough evidence is retained, but becomes sharply unstable once the retained set drops from 22 to 18 checkpoints. This should be interpreted as a diagnostic property of the surrogate under restricted checkpoint information, not as a direct statement about the proprietary scoring engine.

The remediation results applies the surrogate to user specified actions. For supported actions such as SPF and DMARC deployment, the surrogate predicts positive average score impacts, with stronger agreement to trusted reference gains for DMARC than for SPF. At the same time, the remediation reliability analysis shows that these gains are not equally dependable across configurations. The reliability layer plays therefore the role of a screening mechanism.

Several scope limitations remain and point to directions for future work. The empirical study is restricted to the Mail analysis domain of one commercial security rating platform. The remediation analysis currently covers only action families for which we have a reference post-action configuration in our historical dataset. Finally, the CRC guarantees rely on the standard exchangeability assumption between calibration and future cases. In real‑world deployments, however, this assumption may not always hold, particularly under distribution shift or temporal drift..

\paragraph{Ethical and data-handling considerations.}
Our study uses historical security rating records derived from externally observable organization configurations in the Mail analysis domain. We report only aggregated and anonymized results, do not disclose organization identities or sensitive raw records, and present remediation impact predictions as decision support estimates rather than authoritative security recommendations.

 \section{Conclusion}
 \label{sec:conclusion}

In this work, we proposed a surrogate-based approach for remediation score impact prediction in black-box security rating platforms. The approach predicts analysis domain scores without exposing the proprietary scoring engine, studies prediction instability through a controlled checkpoint-restriction diagnostic, and adds a reliability layer calibrated with Conformal Risk Control.

Evaluation on 5,188 real-world Mail organization configurations shows that checkpoint applicability matters for both prediction and reliability. The \textit{applicability aware surrogate} improves score prediction over simpler baselines, while the \textit{sensitivity analysis} reveals a sharp instability transition once the retained checkpoint set becomes too small. The \textit{reliability layer} then uses applicability features to identify unstable cases and to attach selective guarantees to accepted predictions at practical operating points. For \textit{remediation}, the surrogate predicts positive score impacts for supported candidate actions with action-dependent accuracy against trusted reference gains. The remediation oriented reliability analysis shows that predicted instability is useful for identifying cases where predicted gains should be interpreted cautiously.

Overall, the results show that surrogate-based score feedback in opaque security rating settings should not rely on average predictive accuracy alone. Applicability-aware modeling, combined with reliability-aware assessment, enables remediation score impact feedback while preserving scoring engine opacity.

\begin{credits}
\subsubsection{\discintname}
The authors have no competing interests to declare that are relevant to
the content of this article.
\end{credits}

\bibliographystyle{splncs04}
\bibliography{references}

\end{document}